\documentclass[structabstract, letter]{aa} 
\usepackage{graphicx} 
\usepackage[usenames,dvipsnames]{xcolor}
\usepackage{float}
\usepackage{txfonts}
\usepackage{natbib}
\usepackage{units}
\usepackage{url}

\newcommand{\lsi}{LS~I~+61$^{\circ}$303~}

\newcommand{\mwc}{MWC~656}
\newcommand{\agl}{AGL~J2241+4454}

\newcommand{\beq}{\begin{equation}}
\newcommand{\eneq}{\end{equation}}

\newcommand{\rahms}[4]{$#1^{\rm h}#2^{\rm m}#3\mbox{$^{\rm s}\mskip-7.6mu.\,$}#4$} 
\newcommand{\decdms}[4]{$#1^{\circ}#2'#3\mbox{$''\mskip-7.6mu.\,$}#4$} 

\begin{document}

\title{Radio emission from  the Be/black hole binary  MWC 656 }
\author{
S. A.\ Dzib,
M.\ Massi
and 
F.\ Jaron
}

\institute{
 Max-Planck-Institut f\"ur Radioastronomie, Auf dem H\"ugel 69,
 D-53121 Bonn, Germany \\
\email{sdzib, mmassi, fjaron, @mpifr-bonn.mpg.de}
}

   \date{Received 30 June/ Accepted 15 July 2015}
\abstract
{MWC 656 is the recently discovered first binary system case composed of a Be-type star 
and an accreting black hole.
Its low X-ray luminosity indicates that the system is in a quiescent X-ray state.}
{ The aim of our investigation is to establish if
the\ \mwc{} system  has detectable radio
emission and  if the radio characteristics are consistent with those of quiescent black hole systems.}
{We used three archived VLA data sets, one hour each at 3~GHz, and seven new VLA observations, two hours each
at 10~GHz, to produce very high sensitivity images down to $\sim$1$\,\mu$Jy.}
{We detected the source twice in the new observations: in the first VLA run, at apastron passage, 
with a flux density of 14.2$\,\pm\,$2.9  $\mu$Jy and by combining all together the other six VLA runs, 
with  a flux density of  $3.7 \pm 1.4$ $\mu$Jy. The resulting  combined map of the archived  observations
has the sensitivity of $1 \sigma = 6.6\,    \mu Jy$, but no radio emission is detected there.}
{The radio and X-ray luminosities  agree with the behaviour of accreting binary black holes in the hard
and quiescent state. In particular,\ \mwc{} in the $L_X$, $L_R$ plane occupies the same region as 
A0620$-$00  and XTE J1118+480, the faintest  known  black holes up to now.}
{}

\keywords{Radio continuum: stars - X-rays: binaries - X-rays:
  individual (MWC 656) - Gamma-rays: stars}
\titlerunning{The radio source in MWC 656}

\maketitle

\section{Introduction}

\begin{table*}[!th]
\footnotesize
  \begin{center}
  \caption{ MWC\,656 observations and final parameters of maps.}
    \begin{tabular}{ccccccrc}\hline\hline
Epoch  &   Date  & Julian Date& $\Phi$& Synthesized beam  & rms noise & Peak &Flux\,$\pm\,\sigma\,$\tablefootmark{a}\\
   &(day.month\,hh:mm)&          &       &($\theta_{\rm maj}['']\times\theta_{\rm min}[''];$ P.A.[$^\circ$]) & ($\mu$Jy beam$^{-1}$)& ($\mu$Jy bm$^{-1}$)&($\mu$Jy)\\
   \noalign{\smallskip}
\hline\hline\noalign{\smallskip}
\multicolumn{6}{ l }{X-band observations in 2015}\\
   \noalign{\smallskip}
\hline\noalign{\smallskip}
 1 & 22.02\,\,19:15  &2457076.302 & { 0.49}&$1.17\times 1.09;\, 101.7$  & 2.9 & 9.9&14.2\,$\pm$\,2.9\\
 2 & 28.02\,\,19:30  &2457082.313 & { 0.58}&$1.15\times 1.08;\, 104.3$  & 3.8 & ...&$<$11.4\\
 3 & 07.03\,\,22:10  &2457089.424 & { 0.70}&$2.69\times 0.36;\, 58.8$   & 2.3 & ...&$<$6.9\\ 
 4 & 08.03\,\,23:04  &2457090.461 & { 0.72}&$1.62\times 1.06;\, \,84.6$ & 2.4 & ...&$<$7.2\\
 5 & 03.04\,\,19:36  &2457116.317 & { 0.15}&$1.89\times 1.09;\,\,89.3$  & 2.2 & ...&$<$6.6\\
 6 & 10.04\,\,17:01  &2457123.209 & { 0.26}&$1.30\times 0.97;\, -20.6$  & 2.7 & ...&$<$8.1\\  
 7 & 12.04\,\,20:25  &2457125.351 & {0.30}&$1.10\times 1.02;\, -21.9$  & 2.3 & ...&$<$6.9\\
 2 - 7&              &            &     &$0.84\times 0.70;\, -75.1$  & 1.02 & 3.47&3.7\,$\pm$\,1.4\\
 1 - 7&              &            &     &$0.82\times 0.70;\, -75.1$  & 0.95 & 4.15&4.5\,$\pm$\,1.2\\
 \hline\hline\noalign{\smallskip}
\multicolumn{6}{ l }{S-band observations in 2012}\\
   \noalign{\smallskip}
\hline\noalign{\smallskip}
 1 & 05.10\,\,04:38  &2456205.693 & { 0.06}&$0.48\times 0.41;\, 12.3$  & 12.6 & ...&$<$37.8\\
 2 & 15.10\,\,08:03  &2456215.835 & { 0.23}&$0.66\times 0.44;\, 89.6$  & 10.1 & ...&$<$30.3\\
 3 & 06.12\,\,04:54  &2456267.704 & { 0.09}&$0.61\times 0.50;\, 83.0$  & 11.1 & ...&$<$33.3\\ 
 1 - 3 &             &            &     &$0.51\times 0.45;\, 80.5$  &  6.6 & ...&$<$19.8\\
\hline\hline\noalign{\smallskip}

 \label{tab:obsmwc}
\end{tabular}
\end{center}
\tablefoottext{a}{Upper limits at three times the noise level.}
\end{table*}

In 2010 the gamma-ray source\ \agl{}  was detected by the AGILE satellite  
\citep{lucarelli10}. Within the positional error circle of the satellite 
\citet{williams10} found  the  emission-line Be star\ \mwc{}
(also known as HD 215227). 
The first suggested classification for\ \mwc{} was B3 IVne+sh, where $n$ and $e$ indicate broad lines and Balmer emission,
and $sh$ denotes the presence of shell lines, indicated by a sharp central absorption in the line.
The V/R ratio of the $H\gamma$ emission changed significantly in one day, which is unusually fast for most Be stars
and \citet{williams10} pointed out some  resemblance of the spectral properties of
this newly identified gamma-ray
source  to the gamma-ray binary \lsi and its $H\alpha$ variations.
The star\ \mwc{} is at a Galactic latitude of $b=-12^\circ$;
at  a distance of $d=2.6\,\pm\,0.6$ kpc the star results  well below the Galactic plane at $z = -560 \pm 200$ pc.
This is an extreme distance for a normal OB star and \citet{williams10}  suggested
a runaway star, which  reached the current position by a supernova explosion of the companion star.
Timing analysis of variations in the optical photometry, interpreted as orbital modulation,
resulted in 
a period of 60.37$\pm$0.04 days with  the epoch
of maximum brightness at HJD 2453243.3$\pm$1.8.

The binary hypothesis was tested with  a radial velocity study by \citet{casares12}.
Radial velocities folded with the 60.37~d of \citet{williams10} revealed a sine-like modulation.
\citet{casares12}
by the rotational broadening
$ v~\sin{i}\sim~346$~km~s$^{-1}$ estimated the inclination of the orbit  $i >$66$^\circ$ and a  mass of the compact object
in the range $M_c \sim 2.7-5.5$ M$_\odot$.
Moreover, \citet{casares12} found that 
the main parameters of the $H\alpha$ emission line (equivalent width, full width at half maximum, and 
centroid velocity) result to be modulated  by the proposed orbital period.

The turning point in this research came with  the  new optical observations by \citet{casares14}.
To improve the radial velocity curve of the Be star, FeII emission lines from the innermost region
of the Be disk were used instead of stellar broad absorption lines contaminated by the circumstellar
wind emission lines.
On the other hand, the spectra show an HeII 4686 $\r{A}$ double-peaked emission line, indicating a
disk, whose centroid
is modulated with the same 60.37 day period, but in antiphase with respect to 
the radial velocity of the  Be star, as obtained from the FeII line (their Fig. 3).  
This important observation therefore hints at an accretion disk around an invisible companion.
{The orbital solution resulted in $\Phi_{\rm periastron}=$0.01$\pm$0.10 (phase zero set to HJD 2453243.7).}
With new observations, the authors were also able to obtain a better spectral classification for the 
Be star, and determined a B1.5-B2 III classification. Given the mass of this star,  10-16 M$_\odot$
, the implied companion mass is 3.8-6.9 M$_\odot$. This makes MWC 656 the first clear case of a Be-type
star with a black hole companion. The case of \lsi is, in fact, still unclear because of the uncertainties
in the values of mass function and inclination of the orbit \citep{massi04, casares05}.

Observations with XMM-Newton at $\Phi=0.08$ by \citet{munar14} proved that\ \mwc{} is indeed an X-ray binary system.
The X-ray luminosity of 
$\sim 10^{31} {\rm erg\, s}^{-1}$
 points to a stellar mass black hole in quiescence.
The quiescent X-ray state is one of the X-ray states of an accreting black hole  similar to the low-hard
X-ray state.
For a compact object of a few solar masses, the low/hard state
corresponds to $L_{\rm X} \sim 10^{36}$ {${\rm erg\, s}^{-1}$}
but may drop to
$L_{\rm X}= 10^{30.5} - 10^{33.5}$ ${\rm erg\, s}^{-1}$ \citep{mcclintockremillard06}
at its  lowest  phase,   called
the quiescent state. 
The X-ray luminosity in the quiescent and low/hard X-ray states
correspond to a radiatively inefficient, ``jet-dominated'' accretion
mode (Fender et al. 2003).
In this mode, only a negligible fraction of the binding
energy of the accreting gas is directly converted  into radiation.
Most of the accretion power emerges in kinetic
form, as shown for Cygnus X-1  by \citet{gallo05},
and for AGNs through  the relationship between the  Bondi power and the
kinetic luminosity \citep[][and references therein]{merloniheinz07}.
That is, during  the low-hard and quiescent states  the liberated energy of the accretion
is thought to be converted into magnetic energy, which   powers the relativistic jet
observed in these  states \citep{gallo03, fender04, gallo06, smith12}.

The origin of the X-ray emission
during  the low-hard and quiescent states 
is still controversial. The emission may be due to a  Comptonizing corona and/or to the jet \citep[see][]{gallo06}.
Nevertheless,  it is well established that the  X-ray  emission
is related to the radio emission from the jet
 \citep{gallo03, gallo06, merloni03}.
This important nonlinear
scaling between X-ray and radio luminosities, $L_X$, $L_R$, has been demonstrated
to hold, with the addition of a mass term, across the entire black hole mass spectrum, from microquasars to AGN  \citep{merloni03}.

Therefore, if a jet is  associated with\ \mwc,{} its radio emission should be consistent with the 
$L_X$, $L_R$  relationship.
 The aim of our investigation is to establish if the system\ \mwc{} has  an associated radio emission and 
if the radio emission fulfills the $L_X$, $L_R$ relationship.
In this letter, we present new radio observations in the direction of this system.  Section
2 describes the observations and  the  data reduction. In Sect. 3 we report on our results.
Section 4 describes\ \mwc{} in the context of the $L_X$, $L_R$ relationship, and finally,
Sect. 5 gives our conclusions.

\section{Observations}
We obtained seven new X-band (8 to 12 GHz) observations with the Karl G.
Jansky Very Large Array (VLA) of the National Radio Astronomy Observatory
(NRAO) in its B configuration. These observations were made under project
15A-013. The receiver was used in semicontinuum
mode with the 3 bit sampler and 32 different spectral windows (with 
bandwidths of 125 MHz) were recorded simultaneously to cover the full
band. 

Each observing session ran for two hours and was organized as follows. 
First, we spent 4.5 minutes on scans for instrument setups as recommended by 
NRAO\footnote{https://science.nrao.edu/facilities/vla/docs/manuals/obsguide}.
Then, we observed a 5.5 minute scan on the phase calibrator J2255+4202, and the 
large scan was performed to take the slewing time into account. Then we observed
nine cycles of 10.6 minutes on the target followed by 1.0 minutes on the phase
calibrator. We finished the observation with a 5.5 minutes scan in the flux
calibrator 3C147, which was also used as the bandpass calibrator. We spent
$\sim$95 minutes on target per epoch, or a total of $\sim668$ minutes.

The data were edited and calibrated  using the Common Astronomy Software 
Applications (CASA 4.2.2) package, and the VLA calibration pipeline in 
its 1.3.1 version. After calibration images were produced with pixel
sizes of 0.2 arcseconds, a natural weighting, and a multifrequency synthesis
scheme \citep[e.g.,][]{2011A&A...532A..71R}. The noise levels reached for each individual 
observation was about $\sim3~\mu$Jy (see Table \ref{tab:obsmwc}). Additionally,
we produced images from the concatenated UV data from the epoch 2 to 7
and of the seven observations to reach lower noise levels of $1.02~\mu$Jy and $0.95~\mu$Jy,
respectively. These noise levels are in agreement with  the theoretically expected
values.

Finally, we also performed the data reduction of three archived S-band (2-4 GHz) 
data sets taken with the VLA in its A configuration; these are
part of the project 12B-061. Each individual epoch runs for 1.0 hours. The receiver
was also used in semicontinuum mode with the 8 bit sampler and 16 different spectral
windows (with bandwidths of 125 MHz each) were recorded simultaneously to cover the full
band. These observations used the quasar 3C48 as the flux and bandpass calibrator, and quasar
J2202+4216 as the phase calibrator. 

The data were edited, calibrated, and imaged following the same scheme as the new observations.
The resulting sensitivity limits are $\sim$11\,$\mu$Jy and 6.6\,$\mu$Jy, for the maps of individual epochs
and for the map of the concatenated epochs (see also Table \ref{tab:obsmwc}), respectively.
These values are also in agreement with those theoretically expected.
\begin{figure*}[!t]
   \centering
  \includegraphics[height=0.44\textwidth,trim= 10 90 0 100, clip]{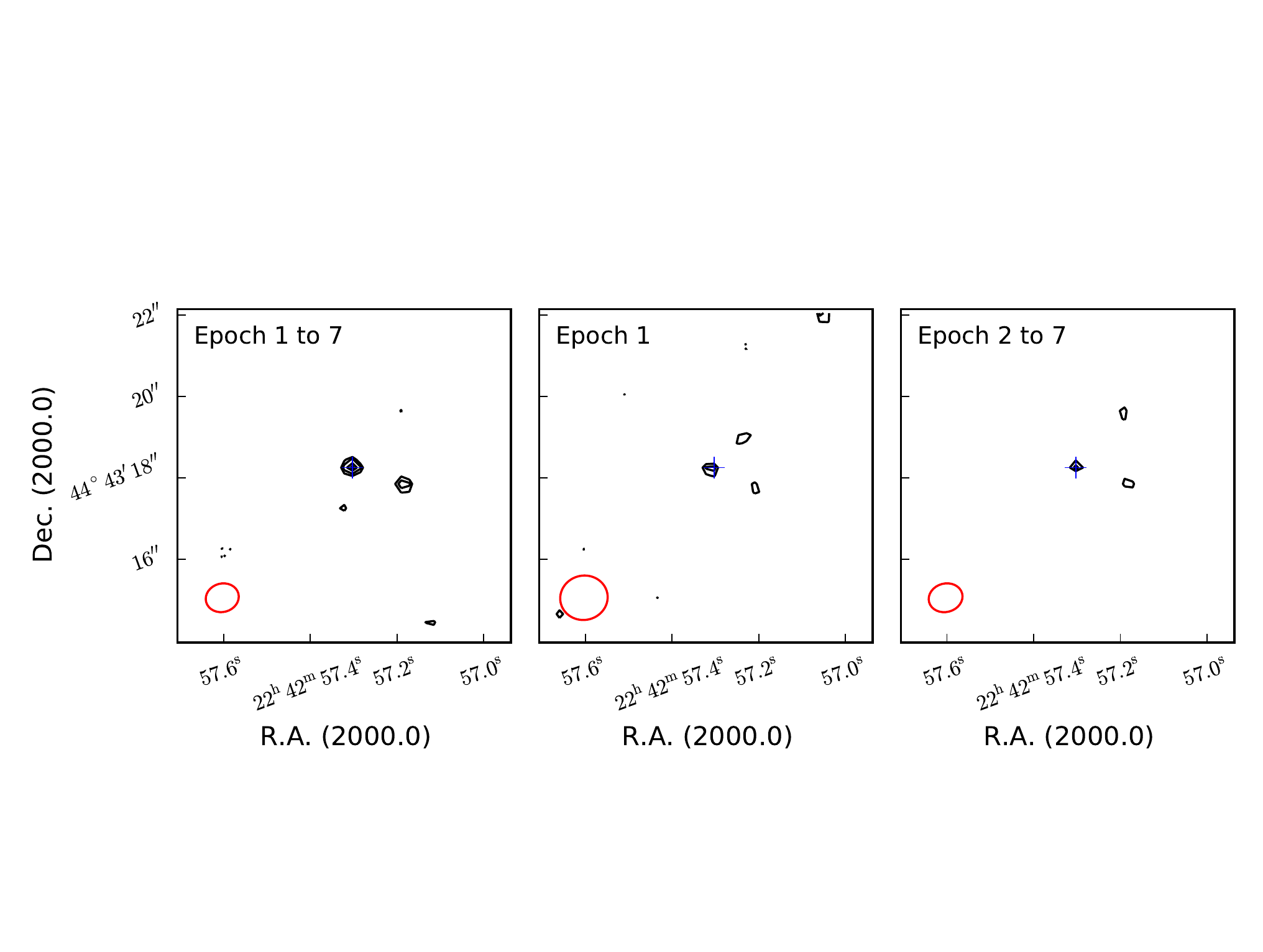}
   \caption{VLA images with radio emission at the MWC 656 position. Left: image of the 
concatenated X-band data from Epoch 1 to 7.
Center:  Epoch~1. Right: image of the concatenated X-band data from Epoch 2 to 6.
   Contour levels are -3, 2.8, 3.2, 3.8, and 4.2 times the noise levels of the image (see Table \ref{tab:obsmwc}).
   The red ellipse at the bottom left is the corresponding synthesized beam. The blue cross indicates
   the optical position of MWC~656.}
   \label{fig:1and6band}
\end{figure*}

\section{Results}
\begin{figure}[!h]
   \centering
  \includegraphics[height=0.60\textwidth,trim= 10 10 15 55, clip]{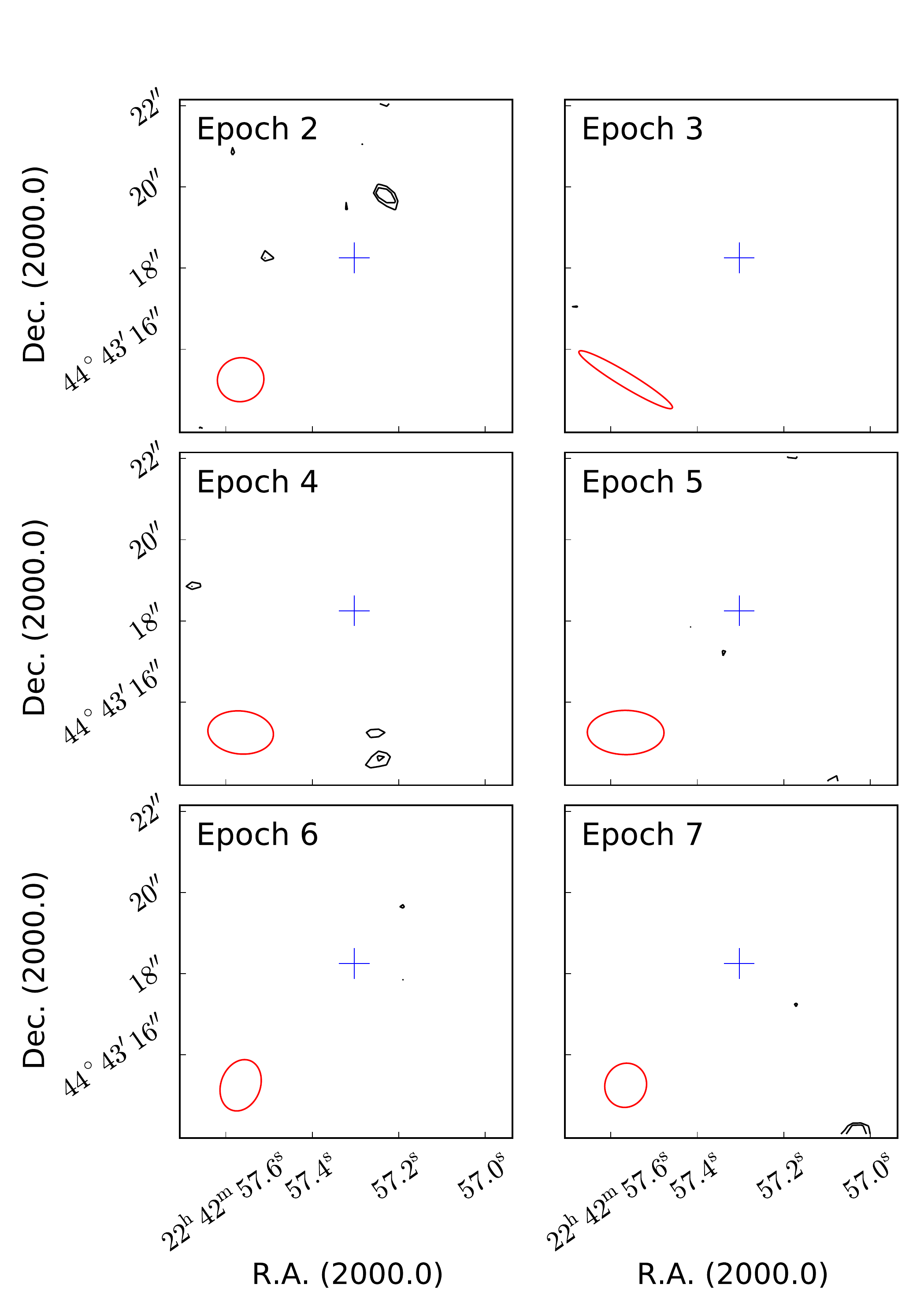}
   \caption{VLA images from epoch 2 to 7. The red circle
   at bottom-left is the synthesized primary beam. Contours are at -3, 2.6,
   and 3 times the noise level. The blue cross at the center represents the position
   of\ \mwc{}. The source remains undetected in each single image. The rms of the images and upper limits
for the flux density of\ \mwc{} are given in Table 1.}
   \label{fig:xband}
\end{figure}
A source with a peak 3.4 times the noise level was detected in the first 
observation at orbital phase $\Phi=0.49$, i.e., at apastron passage.
The source  is coincident with the position of MWC~656 (see
Fig. \ref{fig:1and6band}, center). A Gaussian fit to the source (using {\it imfit} in CASA) yields the flux density of the source to be 14.2\,$\pm$\,2.9 $\mu$Jy.
In the remaining observations, however,
we do not detect any peak above 2.6 times the noise level of the images
(see Fig. \ref{fig:xband}). By adding these six remaining observations,
we detected a source with a peak 3.4 times the noise level, which is coincident
in position with MWC~656 and with the source detected in the first epoch 
(see  Fig. \ref{fig:1and6band}, right). The flux density in this case was 3.7\,$\pm$\,1.4 $\mu$Jy.
The two images of Fig. 1 (center, right) were produced with 
independent data sets, and support our detection of the radio counterpart
of MWC~656.
Interestingly, the flux in the first image (Fig. 1, center) has a flux density almost three times higher 
than that in Fig. 1, right, 
and  this increased flux is occurring   at apastron.
We produce a final map by concatenating the data of the seven observed epochs.
The source is now detected at levels of 4.4 times the noise in the image (Fig. 1, left).
The Gaussian fit to the source  yields the flux density of the source to be 4.5\,$\pm$\,1.2 $\mu$Jy, at the position
RA=\rahms{22}{42}{57}{305} $\pm0\rlap{$^{\rm s}\,$}.005$, 
DEC=\decdms{+44}{43}{18}{23}$\pm$0\rlap{$''$}.\,08,
and is consistent with a point like structure and parameters of Table 1. The source position is in good
agreement with the optical position of MWC~656, 
RA=\rahms{22}{42}{57}{30295}, 
DEC=\decdms{+44}{43}{18}{2525} \citep{2007A&A...474..653V}.

Finally, concerning the archived data (Table 1), we did not detect the source in any  single S-band
epoch, nor in the final image of concatenated observations at levels
above 20$\,\mu$Jy.

\section{MWC 656 in the context of the $L_X$, $L_R$ relationship}

The  radio flux values of  3.7$\,\pm\,$1.4  $\mu$Jy and 14.2$\,\pm\,$2.9  $\mu$Jy at 10~GHz, 
for a distance of 2.6 kpc, and assuming a nearly flat spectrum, corresponds to 
a $L_R$ at 8.6 GHz of $2.6 \times 10^{26}$~erg~s$^{-1}$ and $9.9~\times 10^{26}$~erg~s$^{-1}$, respectively. 
\citet{munar14} determined an  X-ray luminosity of 
$L_X=1.2 \times 10^{31}$ erg s$^{-1}$ from an observation at $\Phi\,=\,0.08$.
We report our new data of\ \mwc{} 
along with the measurements from \citet{corbel13}.
Figure 3 shows (square/black) the radio and X-ray  luminosities for  the  24 
Galactic accreting binary BHs in the hard and quiescence states
reported in Fig. 9 of \citet{corbel13}. 
The position of\ \mwc{},  given in red color,
is rather close to the position of 
  A0620$–$00  and XTE~J1118+480,  which are the weakest quiescent black holes
  known so far \citep{gallo06, gallo14}. 
\begin{figure}[!th]
   \centering
  \includegraphics[]{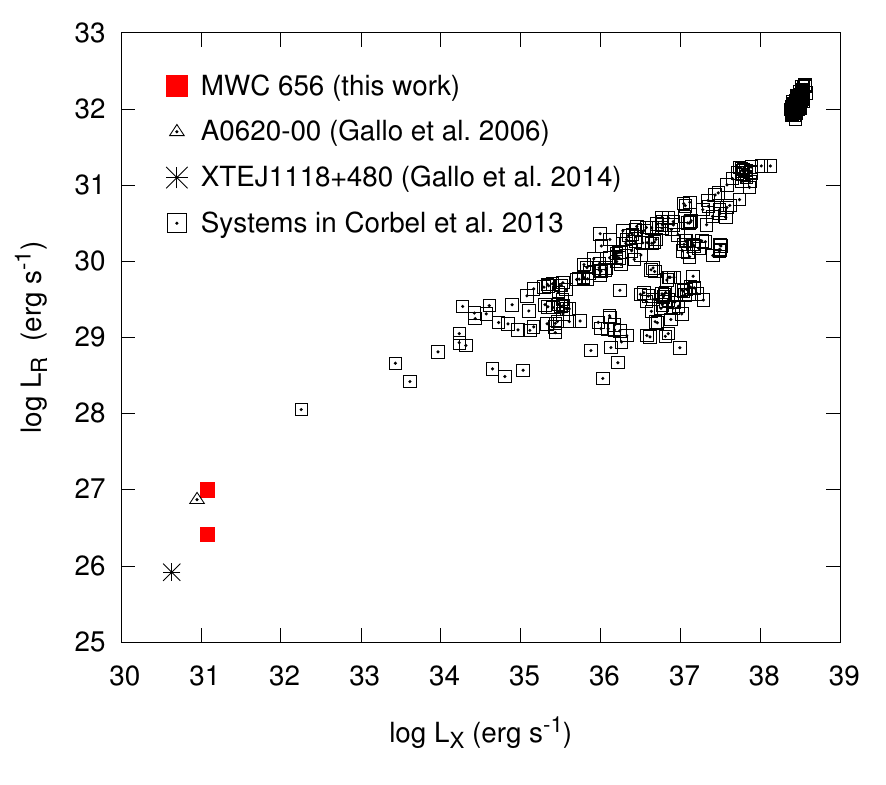}
   \caption{Radio (8.6 GHz) vs X-ray luminosity (1--$10$ keV) diagram. Black squares indicate the positions of the 24 
Galactic accreting binary black holes in the hard and quiescence states, as in Fig. 9  of Corbel et al. (2013; i.e., without the upper limits and neutron stars there present).
The position of MWC 656 is indicated by two  red squares for the two fluxes of 3.7~$\mu$Jy and 14.2~$\mu$Jy.
The position of MWC 656 is rather close to the 
faintest accreting black hole systems known so far, A0620-00 (triangle, \citealt{gallo06}) and 
XTE~J1118+480 (asterisk, \citealt{gallo14}).}
\end{figure}

\section{Conclusions}

Deep VLA observations of the  BH-Be system  MWC 656 have provided us 
with the first radio detection of an accreting  stellar mass black hole with a Be star as companion.
Along with a level of  emission of 3.7$\pm$1.4~$\mu$Jy, determined combining six observations at the different orbital phases  $\Phi=\,$0.58-0.72
and  $\Phi=\,$0.15-0.30, we detected in a single observation around apastron ($\Phi=\,0.49$)
the flux density of 14.2$\,\pm\,$2.9~$\mu$Jy. 

Several works have placed significant constrains on the $L_X$--$L_R$ luminosities of quiescent systems 
\citep[e.g.,][]{gallo06,gallo14,2010MNRAS.409..839C,2011ApJ...739L..18M}.
The nonsimultaneous XMM-Newton X-ray observation by \citet{munar14} at orbital phase $\Phi=0.08$,
also allowed us to test   for\ \mwc\ the X-ray/radio correlation for black holes.
The position of MWC~656 is close to the faintest black holes known so far, which are A0620-00  and XTE~J1118+480.
The secondary star
in A0620--00 is a 0.7 M$_\odot$  K3--K4V star in a 7.75~h orbit around
the black hole \citep{gallo06}.
 The companion star in XTE~J1118+480 is a K5--K8V star with an orbital period of 4.08~h \citep[][and references therein]{gallo14}.
MWC~656  has an orbit of $\sim$ 60~days and the
companion star is a  10-16 M$_\odot$ Be star.
Our observations  support, therefore, the universality of the 
$L_X$, $L_R$ relationship, which  is intimately related to the accretion-ejection coupling 
process,  which seems  to be invariant to different forms of accretion.

\begin{acknowledgements}
We acknowledge Pere Munar-Adrover, Luis F. Rodr\'{\i}guez and Alberto Sanna for comments and suggestions in the manuscript.
The data set of Fig. 3 is provided by S. Corbel with support from the Agence National de la Recherche for the CHAOS project.
The National Radio Astronomy Observatory 
is operated by Associated Universities Inc. under cooperative agreement with the National 
Science Foundation. 
\end{acknowledgements}

\bibliographystyle{aa}

\end{document}